\documentclass[prb]{revtex4}

\usepackage{graphicx}
\usepackage{latexsym}

\usepackage{color}
\newcommand{\beq}{\begin{equation}}
\newcommand{\eeq}{\end{equation}}

\newcommand{\mrm}{\mathrm}
\newcommand{\unit}{\,\mrm}

\newcommand{\Th}{T_\mrm{h}}
\newcommand{\Tha}{T_\mrm{h,1}}

\newcommand{\Tc}{T_\mrm{c}}
\newcommand{\Pc}{P_\mrm{c}}
\newcommand{\rhoc}{\rho_\mrm{c}}

\newcommand{\degreeC}{^\circ\mathrm{C}}

\begin{document} 

\title{Anomalies in bulk supercooled water\\
at negative pressure} 

\author{Ga\"el Pallares$^1$, Mouna El Mekki Azouzi$^1$, Miguel Angel Gonzalez$^2$, Juan Luis Aragones$^2$, Jose Luis F. Abascal$^2$, Chantal Valeriani$^2$ and Fr\'ed\'eric Caupin$^1$}

\affiliation{$^1$Institut Lumi\`ere Mati\`ere, UMR5306 Universit\'e Lyon 1-CNRS, Universit\'e de Lyon, Institut universitaire de France 69622 Villeurbanne Cedex, France}
\affiliation{$^2$Departamento de Quimica Fisica I, Facultad de Ciencias Quimicas, Universidad Complutense, 28040 Madrid, Spain}

\date{27 November 2013}

\begin{abstract}
Water is the most familiar liquid, and arguably the most complex. Anomalies of supercooled water have been measured during decades, and competing interpretations proposed. Yet, a decisive experiment remains elusive, because of unavoidable crystallization into ice. We investigate the state of water that is both supercooled and under mechanical tension, or negative pressure. Liquids under negative pressure can be found in plants or fluid inclusions in minerals. Using such water inclusions in quartz, we report the first measurements on doubly metastable water down to $-15\degreeC$ and around $-100\unit{MPa}$. We observe sound velocity anomalies that can be reproduced quantitatively with molecular dynamics simulations. These results rule out one proposed scenario for water anomalies, and put further constraints on the remaining ones.
\end{abstract}

\maketitle 

\begin{figure}
\centerline{\includegraphics[width=0.65\columnwidth]{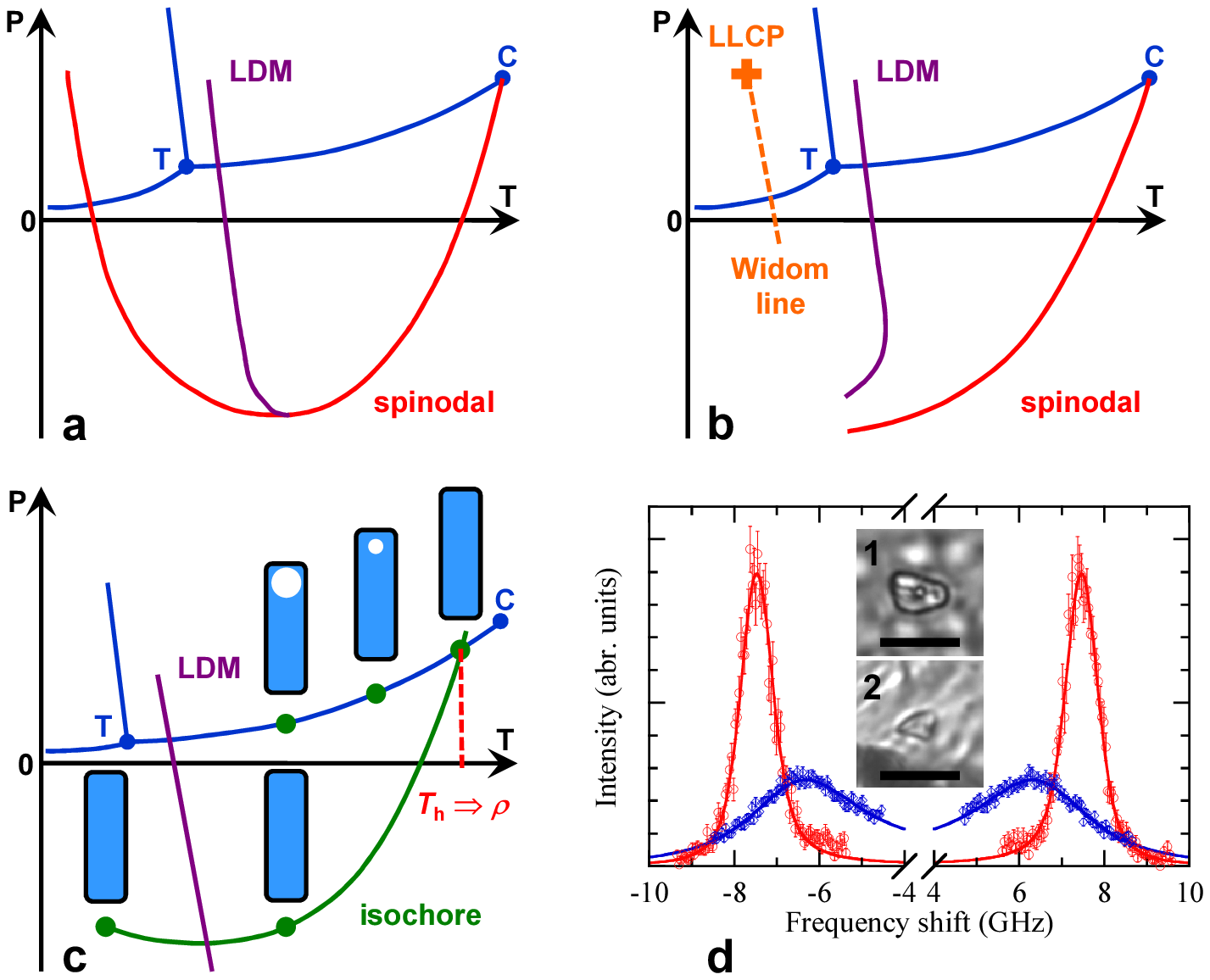}}
\caption{\textbf{Scenarios and experiments for metastable liquid water.} \textbf{a}, Stability-limit conjecture\cite{Speedy_stability-limit_1982}. A schematic phase diagram for water is shown in the pressure-temperature plane with equilibrium transitions between liquid, vapor and ice (blue curves); T is the triple point and C the liquid-vapor critical point. If the LDM reaches the liquid-vapor spinodal at negative pressure, the latter bends to lower tension at lower temperature. This would provide a line of instability at positive pressure responsible for the anomalies observed in supercooled water. \textbf{b}, In the second critical point interpretation\cite{Poole_phase_1992}, the LDM bends to lower temperatures at larger tension, and the spinodal remains monotonic. The anomalies of supercooled water are due to the vicinity of a liquid-liquid critical point (LLCP). Thermodynamic functions exhibit a peak near the Widom line emanating from the LLCP\cite{Sciortino_line_1997,Xu_relation_2005}. \textbf{c}, Berthelot tube method\cite{Zheng_liquids_1991,Alvarenga_elastic_1993,Shmulovich_experimental_2009,El_Mekki_Azouzi_coherent_2013}. A closed, rigid container with a fixed amount of water is heated until the last vapor bubble disappears at $\Th$. Upon cooling, the bubble does not reappear and the liquid follows an isochore (green curve) and is put under mechanical tension. If the density is high enough, cavitation does not occur and the liquid can reach the doubly metastable region, where water is both supercooled and under tension. \textbf{d}, Typical Brillouin spectra. They were obtained with the homogenized sample 1 at $140\degreeC$ (red circles) and $-12\degreeC$ (blue diamonds). The solid curves are fits used to obtain the sound velocity\cite{science_footnote}. Note that the $-12\degreeC$ spectrum was rescaled to the same exposure time as the  $140\degreeC$ spectrum for easier comparison. The samples are shown as inset (scale bars, $10\unit{\mu m}$).\label{fig:general}}
\end{figure}

Water differs in many ways from standard liquids: ice floats on water, and, upon cooling below $4\degreeC$, the liquid density decreases. In the supercooled liquid, many quantities, for example heat capacity and isothermal compressibility, show a large increase. Extrapolation of experimental data suggested a power-law divergence of these quantities at $-45\degreeC$\cite{Speedy_isothermal_1976}. Thirty years ago, the stability-limit conjecture proposed that an instability of the liquid would cause the divergence\cite{Speedy_stability-limit_1982} (Fig.~\ref{fig:general}a). This is supported by equations of state (EoSs), such as the IAPWS EoS\cite{the_international_association_for_the_properties_of_water_and_steam_revised_2009}, fitted on the stable liquid and extrapolated to the metastable regions. Ten years later, the second critical point interpretation, based on simulations\cite{Poole_phase_1992}, proposed that, instead of diverging, the anomalous quantities would reach a peak, near a Widom line\cite{Sciortino_line_1997,Xu_relation_2005} that emanates from a critical point terminating a first order transition between two distinct liquid phases at low temperature (Fig.~\ref{fig:general}b). The two scenarios differ in the shape of the line of density maxima (LDM) of water (see Fig.~\ref{fig:general}a and b). A recent work\cite{El_Mekki_Azouzi_coherent_2013} has added one point on this line at large negative pressure, but this was not enough to decide between the two scenarios. Another, singularity-free scenario, also predicts peaks instead of divergence, but without a liquid-liquid transition (LLT)\cite{Sastry_singularity-free_1996}. Up to now, despite tremendous efforts\cite{Debenedetti_supercooled_2003-1}, experiments have failed to detect a LLT or peaks in bulk supercooled water: crystallization always occurred before any extremum is reached.

Measurements on a metastable liquid are difficult to perform. In bulk water at positive pressure, decisive experiments to discriminate between the proposed scenarios have been precluded by unavoidable crystallization. To circumvent this problem, water proxies have been used: water confined in narrow pores\cite{Liu_observation_2007}, or bulk water-glycerol mixtures\cite{Murata_liquidliquid_2012}. Although the results supported the second critical point interpretation, their relevance to bulk water is not straightforward.

Here we study bulk water samples, a few microns in diameter, in the doubly metastable region: the liquid is simultaneously supercooled and exposed to mechanical tension or negative pressure. Negative pressures occur in nature, e.g. in the sap off trees, under the tentacles of octopi, or in fluid inclusions in minerals\cite{Caupin_cavitation_2006,Caupin_stability_2013}. The study of the largest tensions achievable in water was pioneered by the group of Angell\cite{Zheng_liquids_1991}. They used a `Berthelot tube' technique (Fig.~\ref{fig:general}c), based on isochoric cooling of a micrometer size inclusion of water in quartz. Tensions as large as $-140\unit{MPa}$ have been reported, and confirmed by others\cite{Alvarenga_elastic_1993,Shmulovich_experimental_2009,El_Mekki_Azouzi_coherent_2013}, which exceed by far the limit of other techniques\cite{Caupin_cavitation_2006,Caupin_stability_2013}. It was already recognized in the work of Angell that the high density water inclusions that were able to survive cooling to room temperature without cavitation were also able to be supercooled below $0\degreeC$. Indeed, when the isochore crosses the line of density maxima of water, the tension is released and cavitation becomes less likely. Another study\cite{Henderson_Berthelot-Bourdon_1980} using macroscopic Berthelot tubes also reached the doubly metastable region, but the tensions were around $-10\unit{MPa}$ only.

In order to reach large tensions, we use two microscopic inclusions of water in quartz (Fig.~\ref{fig:general}d, inset)\cite{science_footnote}. We perform Brillouin light scattering experiments on these samples; this technique gives access to the sound velocity within the liquid\cite{science_footnote}. Several Brillouin light scattering studies on supercooled water at ambient pressure are available\cite{Teixeira_Brillouin_1978,Conde_analysis_1980,Conde_analysis_1982,Maisano_evidence_1984,Magazu_relaxation_1989,Cunsolo_velocity_1996}, but only one work investigated water under tension\cite{Alvarenga_elastic_1993}: all samples in that study cavitated above room temperature except one with a density close to that of our sample 1. However, measurements in Ref.~\citenum{Alvarenga_elastic_1993} were reported only down to $0\degreeC$, and the direct comparison to an extrapolated EoS was not considered. Our work extends the covered range to supercooled water under tension, 
reporting measurements down to $-15\degreeC$ along two isochores at $\rho_1=933.2 \pm 0.4 \unit{kg\,m^{-3}}$ and $\rho_2=952.5 \pm 1.5 \unit{kg\,m^{-3}}$, and reaching pressures beyond $-100\unit{MPa}$ (supplementary online text). Representative spectra are shown in Fig.~\ref{fig:general}d. Such spectra are analyzed to give the zero frequency sound velocity $c$\cite{science_footnote}. To support the experimental results, we also perform molecular dynamics simulations of $c$ with TIP4P/2005 water\cite{Abascal_general_2005} at the same thermodynamic conditions\cite{science_footnote}.

\begin{figure}
\centerline{\includegraphics[width=0.65\columnwidth]{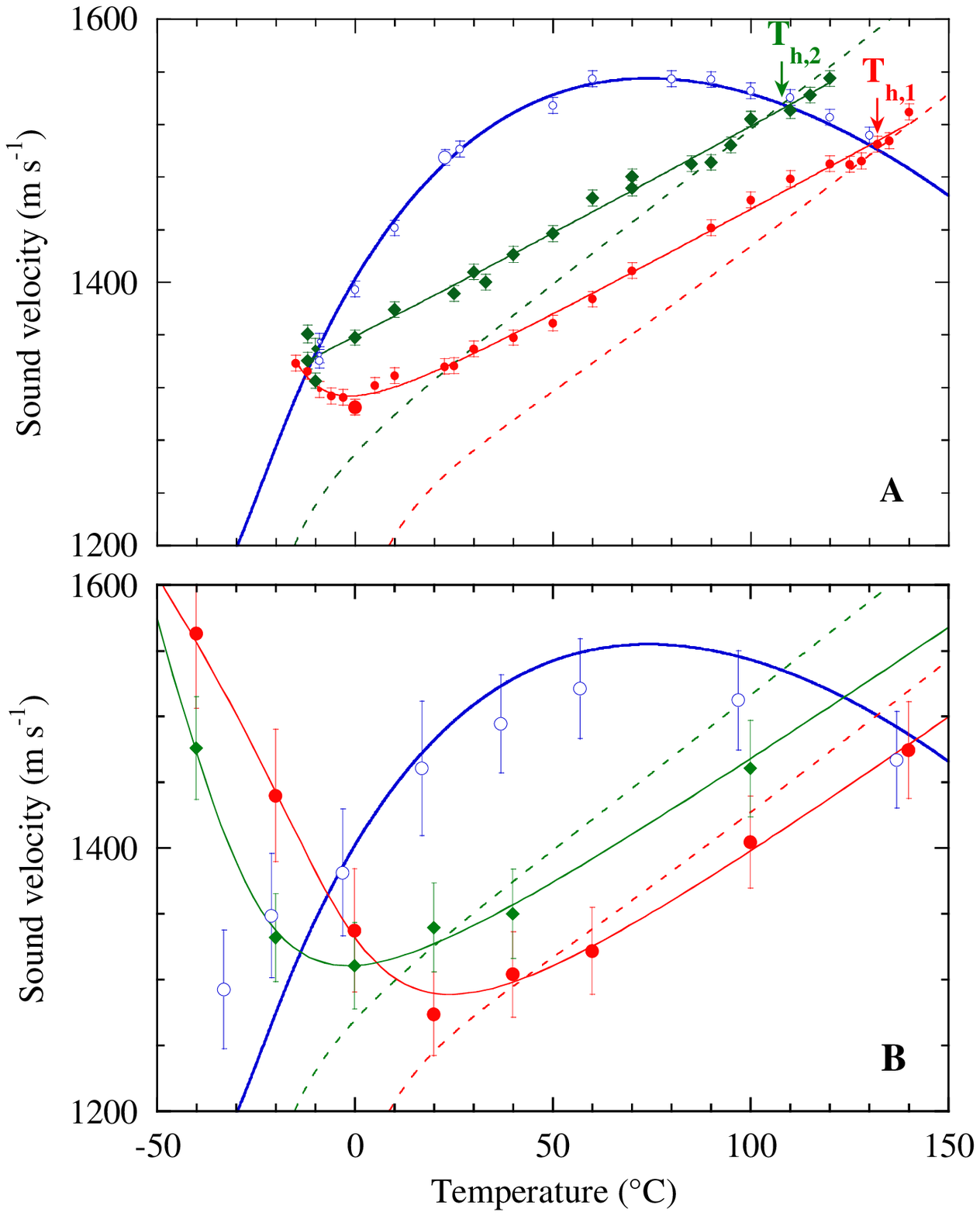}}
\caption{\textbf{Sound velocity as a function of temperature.} 
The IAPWS EoS is used to plot the sound velocity along the binodal (thick blue curve) and along the isochores at $\rho_1$ (dashed red curve) and $\rho_2$ (dashed green curve). \textbf{A}, comparison with experiments. The symbols show our measurements on sample 1 after cavitation (open blue circles) and on metastable samples 1 (filled red circles) and 2 (filled green diamonds). The three symbol sizes correspond to three pinhole sizes on the spectrometer. The solid red and green curves are guides to the eye. The arrows show the homogeneization temperatures of samples 1 and 2 as observed under the microscope. \textbf{B}, comparison with simulations of TIP4P/2005 water. The sound velocity was calculated along the TIP4P/2005 binodal (open blue circles), and the isochores at $\rho_1$ (filled red circles) and $\rho_2$ (filled green diamonds). The solid red and green curves are guides to the eye. Whereas the IAPWS EoS predicts a monotonic variation of $c$ along the isochores, both experiments and simulations find that $c$ reaches a minimum and increases above the values on the binodal at low temperature.\label{fig:c-T}}
\end{figure}

Figure~\ref{fig:c-T} shows the experimentally measured and the numerically computed values of $c$ as a function of temperature at several thermodynamic conditions. Let us first describe the measurements on sample 1 after cavitation, at temperatures up to $\Tha$: at these conditions the liquid is in equilibrium with its vapor. The measured sound velocity is in excellent agreement with the known sound velocity along the binodal and with our simulations of TIP4P/2005 water along its binodal\cite{Vega_vapor-liquid_2006}. The agreement between simulations and tabulated experimental data also illustrates the quality of the potential used to simulate water\cite{Abascal_general_2005}. When the temperature of sample 1 reaches $\Tha$, the last vapor bubble disappears leaving the inclusion entirely filled with liquid at density $\rho_1$. Upon further heating, the pressure increases along the $\rho_1$ isochore. Once more, the measurements agree with the sound velocity from the known EoS and from our simulations along the $\rho_1$ isochore. Note that the measurements leave the binodal exactly at $\Tha$, which has been determined independently by direct observation under the microscope. These consistency further corroborates the robustness of our data. 

\begin{figure}
\centerline{\includegraphics[width=0.65\columnwidth]{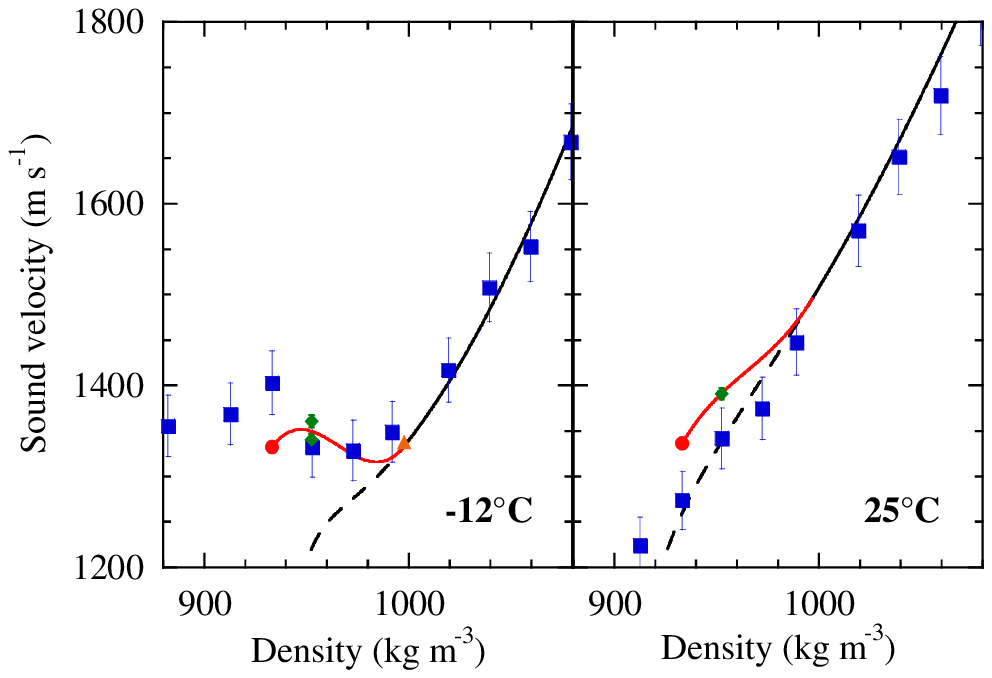}}
\caption{\textbf{Sound velocity versus density at $-12\degreeC$ and $25\degreeC$.} The IAPWS EoS is shown in black, as a solid curve above the binodal, and as a dashed curve for its extrapolation to lower density. Our measurements for $\rho_1$ and $\rho_2$ are shown as filled red circles and green diamonds, respectively. The red curves are guides to the eye chosen to connect the data to the IAPWS EoS above the binodal with the correct slope. Note that the IAPWS EoS reproduces accurately experimental data above the binodal for stable water (e.g. at $25\degreeC$), and data along the binodal for supercooled water (e.g. at $-12\degreeC$). To illustrate the latter, we have included in the left panel the experimental value of the sound velocity at $-11.2\degreeC$ (orange triangle) from ref.~\citenum{Magazu_relaxation_1989} (see SOM text). The solid blue squares give the results of the simulations of TIP4P/2005 water; note that the simulations in the right panel were obtained at $20\degreeC$. At high temperature, measurements and simulations agree with the IAPWS EoS, whereas at low temperature, they suggest the occurence of a minimum and maximum which are absent from the extrapolation of the IAPWS EoS.\label{fig:c-rho}}
\end{figure}

Next, we make sample 1 metastable by cooling along the $\rho_1$ isochore. We observe that below $\Tha$, the measured sound velocity starts diverging with respect to the one extrapolated using the IAPWS EoS. In contrast, it agrees with our simulations. The IAPWS EoS predicts a re-entrant liquid-vapor spinodal: in this case, one would expect the sound velocity curve to reach a small value when the isochore approaches the spinodal at low temperature (for $\rho_1$, the IAPWS EoS predicts that they meet at $-33.5\degreeC$ with $c=829\unit{m\,s^{-1}}$). In contrast, both experiments and simulations give a sound velocity that reaches a minimum near $0\degreeC$, before increasing on further cooling. Therefore, the isochore at $\rho_1$ does not approach the liquid-vapor spinodal as expected from the IAPWS EoS. Since we follow an isochore, the sound velocity minimum corresponds to a maximum in the adiabatic compressibility. Eventually, the sound velocity reaches a value \textit{higher} than the value that one would expect along the binodal at $-15 \degreeC$, even though $\rho_1$ lies \textit{below} the liquid density on the binodal, $\rho_0=996.3\unit{kg\,m^{-3}}$ at $-15\degreeC$ (ref.~\citenum{Hare_density_1987}).

To confirm our results, we have repeated measurements and simulations on sample 2 with a higher density $\rho_2$. We find a similar deviation from the extrapolation of the IAPWS EoS. In the simulations, the sound velocity reaches a minimum, which is not clear in the experiments that seem to reach a plateau. This is consistent with the fact that the simulations find the minimum for $\rho_2$ at a temperature lower than for $\rho_1$, while for $\rho_1$ the experiment finds the minimum at a temperature lower than the simulations. Therefore it is likely that the minimum for $\rho_2$ lies at temperatures below the one reached in the experiment. In both experiment and simulations, while at $-12 \degreeC$ $\rho_2$ is \textit{between} $\rho_1$ and $\rho_0$, the corresponding sound velocity is even slightly \textit{higher} than both. Therefore, the sound velocity must reach a minimum between $\rho_2$ and $\rho_0$ along the $-12\degreeC$ isotherm. This is more clearly seen on Fig.~\ref{fig:c-rho}: the sound velocity at $\rho_0$, $\rho_1$ and $\rho_2$ virtually fall on a horizontal line, but the slope of $c(\rho)$ at $\rho_0$ necessarily implies a minimum. This observation agrees with the predictions of the simulations but contrasts those of the IAPWS EoS.

To explain the observed anomalies, it is interesting to look for other systems with a minimum in $c(\rho)$ along an isotherm. This actually occurs in all fluids, in their supercritical phase: the sound velocity at a temperature above the liquid-vapor critical temperature passes through a minimum when the density crosses the critical density (see for instance Fig.~8 of Ref.~\citenum{Ohmori_thermal_2001} for methanol and ethanol). Based on this observation, we propose that the anomalies of the sound velocity, observed both in experiments and in simulations, are a signature of supercritical phenomena. Here, in addition, the sound velocity reaches a maximum when the density decreases (Fig.~\ref{fig:c-rho}). This is because eventually the liquid-vapor spinodal density has to be reached, and the sound velocity has then to become small. The simulations can then be used as a guide to better understand the origin of such anomalies. Several potentials predict a LLT in the supercooled region, ending at a LLCP. Based on the locus of maxima of $\kappa_T$ along isobars, a LLCP for TI4P/2005 has been proposed at $\Tc=193\unit{K}$, $\Pc=135\unit{MPa}$, and $\rhoc=1012\unit{kg\,m^{-3}}$\cite{Abascal_Widom_2010}. Water above $193\unit{K}$ would thus be in a supercritical state associated to the LLT.

\begin{figure}
\centerline{\includegraphics[width=0.65\columnwidth]{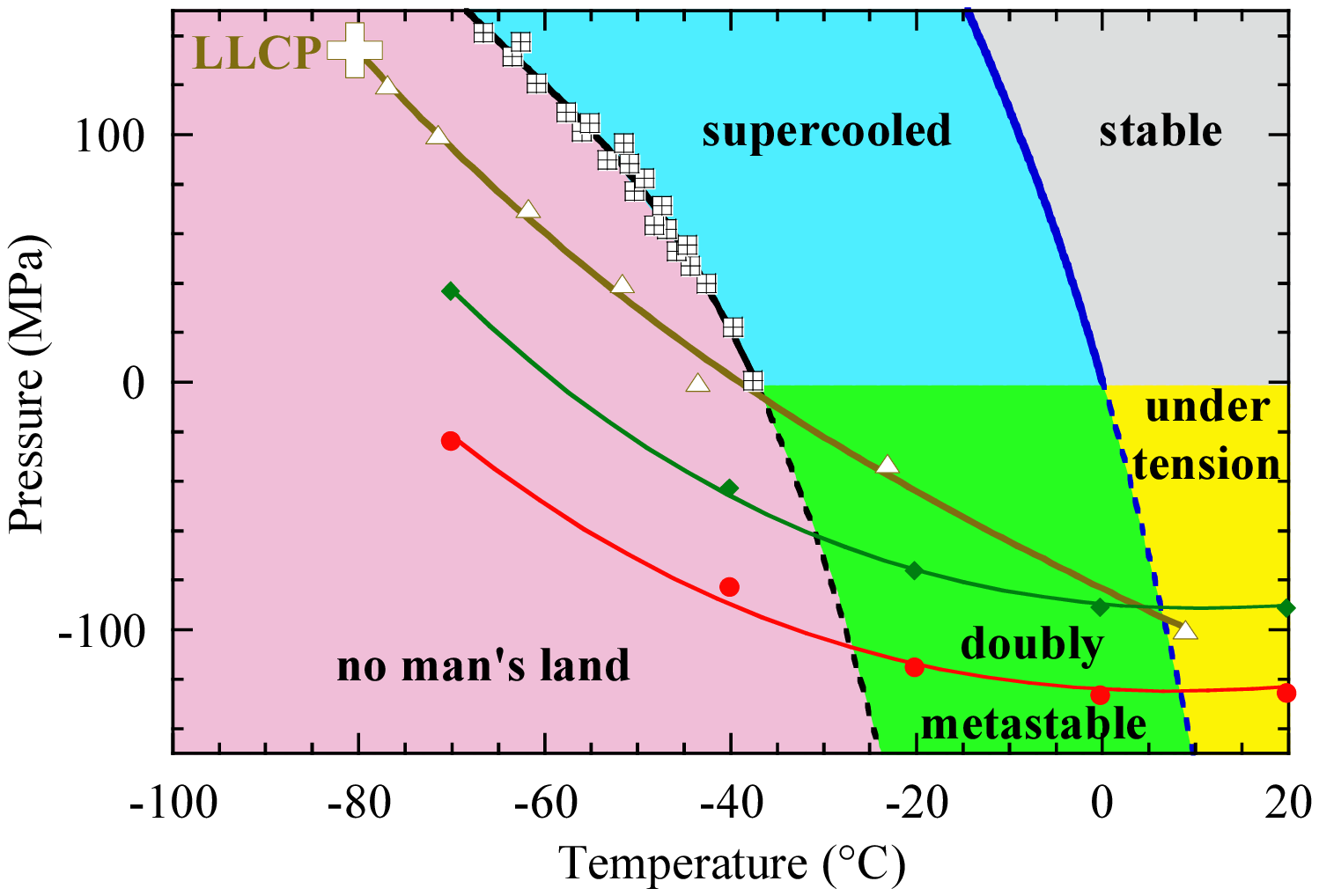}}
\caption{\textbf{Pressure-temperature phase diagram of water.} Colored areas are used to identify the different possible states for liquid water. The melting line of ice Ih is shown at positive pressure by a solid blue curve and its extrapolation to negative pressure by a dashed blue curve. The black crossed squares show the experimental supercooling limit\cite{Kanno_supercooling_1975}. They define the experimental homogeneous nucleation line (solid black curve), which is extrapolated here to negative pressure (dashed black curve). The $\rho_1$ and $\rho_2$ isochores of TIP4P/2005 water are shown by the red circles and curve, and green diamonds and curve, respectively. Simulations of TIP4P/2005 water are performed to find the maximum $\kappa_T$ along several isobars (white triangles), defining the LM$\kappa_T$ (brown curve), that might emanate from a liquid-liquid critical point (LLCP, white plus symbol). The LM$\kappa_T$ becomes accessible to experiments in the doubly metastable region only.\label{fig:Widom}}
\end{figure}

One may wonder why all previous experiments at positive pressure failed to detect a peak in thermodynamic functions. Simulations with a LLT show that specific quantities (such as isothermal compressibility $\kappa_T$ or isobaric heat capacity) reach a peak on different lines, but that these lines all come close to each other and to the Widom line near the LLCP\cite{Xu_relation_2005}. The locus of maxima for $\kappa_T$ along isobars (LM$\kappa_T$) has been recently computed for TIP4P/2005 water at positive pressure\cite{Abascal_Widom_2010}. We have now computed this line also at negative pressure (supplementary online text). The results are shown on Fig.~\ref{fig:Widom} and compared to the isochores we studied and to the experimental line of homogeneous crystallization\cite{Kanno_supercooling_1975}. At positive pressure, the LM$\kappa_T$ is not accessible to experiments since it lies in the so-called `no man's land'\cite{Mishima_relationship_1998}, a region where bulk liquid water cannot be observed experimentally. However, at negative pressure, the slope of the LM$\kappa_T$ becomes less negative than that of the line of homogeneous crystallization. Since the latter keeps the same slope, the LM$\kappa_T$ line leaves the `no man's land' and enters the doubly metastable region that we have now shown is accessible to quantitative experimentation.

Is there a way to directly observe the LLT? If it exists, it is likely to lie in the `no man's land'. Therefore it will be hard to decide between the second critical point interpretation and the singularity free scenario\cite{Sastry_singularity-free_1996}; note however that the latter can be seen\cite{Stokely_effect_2010} as a LLT with a critical point at $0\unit{K}$. What our results do show is that the liquid-vapor spinodal does not reach the location predicted by extrapolation from the IAPWS EoS, and that the measured sound velocity quantitatively agrees with the one obtained with simulations of TIP4P/2005 water, which predicts a LLT. The doubly metastable region therefore appears like a promising experimental territory to test other predictions from the models proposed to explain water anomalies.


\begin{acknowledgments}
\textbf{Acknowledgements} We thank C. Austen Angell and Jos\'e Teixeira for discussions; V\'eronique Gardien for providing us with sample 2; and Abraham D. Stroock and Carlos Vega for suggestions to improve the manuscript. The team at Lyon acknowledges funding by the ERC under the European Community's FP7 Grant Agreement 240113, and by the Agence Nationale de la Recherche ANR Grant 09-BLAN-0404-01. C.V. acknowledges financial support from a Marie Curie Integration Grant PCIG-GA-2011-303941 ANISOKINEQ and thanks the Ministerio de Educacion y Ciencia and the Universidad Complutense de Madrid for a Juan de la Cierva fellowship. The team at Madrid acknowledges funding from the MCINN Grant FIS2010-16159.

\textbf{Supplementary Materials} Materials and Methods. SOM Text. Figs. S1 to S7. Table S1 to S3.
 
\end{acknowledgments}



\begin{thebibliography}{10}
\bibitem{science_footnote}
\protect{Materials and} methods are available as supporting material.
\end{thebibliography}
\end{document}